\documentclass{article}

\usepackage[preprint]{nips_2018}

\usepackage[utf8]{inputenc} 
\usepackage[T1]{fontenc}    
\usepackage{hyperref}       
\usepackage{url}            
\usepackage{booktabs}       
\usepackage{amsfonts}       
\usepackage{nicefrac}       
\usepackage{microtype}      
\usepackage{amsmath}
\usepackage{amssymb}
\usepackage{graphicx}
\usepackage{subcaption}
\usepackage{amsthm}
\usepackage[english]{babel}
\usepackage[toc,page]{appendix}
\usepackage{thm-restate}
\usepackage{cleveref}
\usepackage{color}

\newcommand{\norm}[1]{\left\lVert#1\right\rVert}

\title{Accelerating Flash Calculation through Deep Learning Methods}

\author{
  Yu Li \\
  KAUST \\
  CEMSE\\
  CBRC\\
  \And
  Tao Zhang \\
  KAUST \\
  PSE\\
  CTPL\\
  \And
  Shuyu Sun $^{*}$ \\
  KAUST \\
  PSE\\
  CTPL
  \And
  Xin Gao \thanks{All correspondence should be addressed to Xin Gao (xin.gao@kaust.edu.sa) and Shuyu Sun (shuyu.sun@kaust.edu.sa).} \\
  KAUST \\
  CEMSE\\
  CBRC\\
}

\begin{document}

\maketitle

\begin{abstract}

In the past two decades, researchers have made remarkable progress in accelerating flash calculation, which is very useful in a variety of engineering processes. In this paper, general phase splitting problem statements and flash calculation procedures using the Successive Substitution Method are reviewed, while the main shortages are pointed out. Two acceleration methods, Newton's method and the Sparse Grids Method are presented afterwards as a comparison with the deep learning model proposed in this paper. A detailed introduction from artificial neural networks to deep learning methods is provided here with the authors' own remarks. Factors in the deep learning model are investigated to show their effect on the final result. A selected model based on that has been used in a flash calculation predictor with comparison with other methods mentioned above. It is shown that results from the optimized deep learning model meet the experimental data well with the shortest CPU time. More comparison with experimental data has been conducted to show the robustness of our model. 

\end{abstract}
\section{Introduction\label{sec:Introduction}}

Vapor-liquid equilibrium (VLE) is of essential importance in modeling the multiphase and multicomponent flow simulation for a number of engineering processes \cite{Sun:2007aa, Sun:2005aa}. Knowledge of the equilibrium conditions in mixtures can be obtained from data collected in direct experiment or using thermodynamic models, including activity coefficients at low system pressure or fugacity coefficients at high system pressure \cite{Prausnitz:1998aa}. In the last two decades, the application of equilibrium calculation using equations of state (EOS) that describes mixing rules with experience coefficients has been proposed and widely discussed \cite{Iliuta:2000aa, Tan:2004aa, Tan:2005aa}. A realistic EOS, e.g., Peng-Robinson (PR EOS), is generally considered as an appropriate thermodynamic model to correlate and predict vapor\textendash liquid equilibrium (VLE) conditions, due to the long-time improvement developed with applications in different aspects. The calculation procedures using EOS have been extensively studied and modified. The EOS parameters, describing concentration, combination and interaction between binary mixtures, can decide the accuracy in correlating the VLE process. In practice, such parameters are generally obtained by fitting experimental data under the temperature at which VLE is required. However, in most EOS calculations, iterations are needed, which makes it less suitable for time sensitive applications. For cases like phase equilibrium in underground heated flow, as the concentration of many possible components varies greatly in a wide range, it is difficult to correlate them and predict the parameters exactly from experiments. Some very small concentrations of components are essential parts, which make the parameter calculation more difficult. 

To speed up flash calculation, attentions have been paid to other numerical tools, such as sparse grids technology and artificial neural networks (ANN) \cite{Iliuta:2000aa,Wu:2015aa,Azari:2013aa,Wu:2015ab,Nguyen:2007aa,Nikkholgh:2010aa}. Sparse grids technology is often considered preferable in coupling with flow as the surrogate model created in the offline phase can be used repeatedly \cite{Wu:2015ab}. However, the generation of the surrogate model is still time-consuming. {\color{black} On the other hand, due to its ability to capture the relationship among large amount of variables, especially for non-linear correlations, ANN has attracted considerable attention for performing acceleration. It has been reported that ANN has been used in the thermodynamic properties calculation successfully, including vompressibility factor, vapour pressure and viscosity. For example, ANN has shown the efficiency, as well as the ability to estimate the shape factors and density of regregerants, which is a function of temperature, and then been extended to solve the corresponding state model} \cite{Nguyen:2007aa, Azari:2013aa,Nikkholgh:2010aa}.

Since the breakthrough of AlexNet \citep{RN69} in 2012, deep learning has made profound impact on both the industry and the academia \citep{RN4}. Not only has it revolutionized the computer vision field, improving the performance of image recognition and object detection dramatically \citep{RN7,renNIPS15fasterrcnn}, and the natural language processing field \citep{RN215}, setting new record for speed recognition and language translation, but it has also enabled machines to reach human level intelligence in some certain tasks, such as the Go game \citep{RN16}. In addition to those well-known tasks that deep learning is especially expert in, deep learning has been applied to a broad range of problems, such as protein binding affinity prediction \citep{RN141}, enzyme function prediction \citep{RN140}, structure super-resolution reconstruction \citep{DLBI}, the third generation sequencing modeling \citep{deepsimulator}, particle accelerator data analysis \citep{RN287} and modeling brain circuits \citep{RN288}. 
Such a great potential of deep learning comes from its significant performance improvement over the traditional machine learning algorithms, such as support vector machine (SVM). The traditional machine learning methods usually consider just one layer non-linear combination of the input features while the deep learning method can consider ultra-complex non-linear combination of the input features by taking advantage of multiple hidden layers. During training, the back propagation algorithm increases the weight of the feature combination which is useful for the final classification or regression problem to emphasize the useful features while decreases the weight of those unrelated feature combinations. In spite of the universal approximation theorem, which states that we can approximate any continuous functions using a feed-forward network with a single hidden layer containing a finite number of neurons, the success of deep learning shows the potential of fitting VLE using multi-layer neural networks.


{\color{black} In this paper, binary components flash calculation accelerations are studied. We first introduce the main concept of flash calculation and general problem statement. After that, we explain a popular method in flash calculation, Successive Substitute Method, in details as well as the main shortage of this method. Some suggestions are presented based on our experience to help reach the equilibrium state. Afterwards, two optimization methods, namely, the Newton's method and the Sparse Grids Method, are briefly introduced. Deep learning methods based on artificial neural networks are presented in the following section.  Our deep learning model is optimized based on the investigation of its testing behaviors with different factors and features. Then, all the four methods are applied in flash calculation test, and we show the comparison of their results and CPU time needed. The fastest deep learning method is finally tested with various flash calculation problems, and compared with the experimental data and results from classical SSM. Remarks are made and conclusions are drawn based on the results and comparisons.}

\section{Flash Calculation Methods}

\subsection{General Phase Splitting Problem Statement}

{\color{black} In two-phase compositional flow, component compositions change with the thermodynamic conditions, and that is what flash calculations determine. Phase splitting may occur with the changing of temperature or pressure, which makes the composition determination more difficult. Wider properties at phase equilibrium statements can also be predicted through flash calculation, and the modeling and simulation have been widely reported in literatures \cite{li2017numerical,nghiem1982general}. A general phase splitting problem is often defined with an assumption of a thermodynamic equilibrium state among the components. 
Common phase splitting problems include the NPT flash $(T,P,x,y)$ and the NVT flash$(T,V,x,y)$. For both types, temperature and pressure remain the same for all the phases, while other properties need to be calculated based on the \textquotedblleft thermodynamic equilibrium condition\textquotedblright. Fugacity is usually selected to express the equilibrium condition of each component in engineering application, and NPT flash is more popular compared with NVT flash. }

In a typical NPT phase splitting problem, we are given the following data as our input quantities: the moles of feed $F$ (no need if interested in extensive quantities only); feed composition $z_{i},i=1,\cdots,M$; temperature $T$, and pressure $P$ and physical properties of components such as critical temperature $T_{c}$ and critical pressure $P_{c}$. The purpose is to determine the vapor-phase composition $y_{i},i=1,\cdots,M$, and liquid-phase composition $x_{i},i=1,\cdots,M$.
Popular solutions include several different variants of the successive substitution method (SSM). {\color{black} Molar densities, mass densities and isothermal compressibility of each phase can be determined further based on the SSM result.  Finally, the saturation of the flow, the partial molar volume of each species and the two-phase isothermal compressibility of the flow can be calculated. }


\subsection{Successive Substitution Method}

The Successive Substitution Method (SSM) is also known as Successive Substitution Iteration (SSI), which is mainly the updating of vapor-liquid equilibrium ratio,  $K_i$,

 \begin{align}
 y_i = K_i x_i,\,\, i=1,2,\cdots,M,
 \end{align}
also known as the phase equilibrium constant, the partition constant or the distribution coefficieint, and can also be defined from 
 
 \begin{align}
 K_i = K_i(T,p,x_1,\cdots,x_M,y_1,\cdots,y_M) = \frac{\varphi_i^L(T,p,x_1,x_2,\cdots,x_M)}{\varphi_i^V(T,p,y_1,y_2,\cdots,y_M)}.
 \end{align}
 
The equilibrium condition is equivalent to $p y_i = K_i x_i$. The fugacity can be computed by PR-EOS as  

 \begin{align}
\ln\varphi^L_{i}=\dfrac{b^L_{i}}{b^L}\left(Z^L-1\right)-\ln\left(Z^L-B^L\right) - \frac{A^L}{2\sqrt{2}B^L}\left(\frac{2\sum_{j=1}^{M}x_{j}a^L_{ij}}{a^L}-\frac{b^L_{i}}{b^L}\right)\ln\frac{Z^L+2.414B^L}{Z^L-0.414B^L},
 \end{align}
 
 \begin{align}
\ln\varphi^V_{i}=\frac{b^V_{i}}{b^V}\left(Z^V-1\right)-\ln\left(Z^V-B^V\right)- \frac{A^V}{2\sqrt{2}B^V}\left(\frac{2\sum_{j=1}^{M}y_{j}a^V_{ij}}{a^V}-\frac{b^V_{i}}{b^V}\right)\ln\frac{Z^V+2.414B^V}{Z^V-0.414B^V}.
 \end{align}
 
 If the K-values are given, we can solve the Rachford-Rice equation for $\beta$:
 
 \begin{align}
 \sum_{i=1}^M \frac{(K_i-1) z_i}{1+\beta(K_i-1)}=0.
 \end{align}

The coefficient $\beta$ is more convinced in the composition computation compared with K-values.

{\color{black} The detailed algorithm in SSM could be summarized into 5 steps:

Step 1. Prepare the parameters, including $a$ and $b$, based on the critical pressures and temperatures of each component as well as their acentric factors.  

\begin{align}
a(T_c)={\frac {0.45724\,R^{2}\,T_{c}^{2}}{p_{c}}}, \,\,\, b=b(T_c)={\frac {0.07780\,R\,T_{c}}{p_{c}}},
a(T) = a(T_c) \left(1 + m \left(1-T_r^{0.5}\right)\right)^2,
\end{align}
where m is calculated by 
\begin{align}
m = 0.37464 + 1.54226 \omega - 0.26992 \omega^2, for 0<\omega<0.5,
\end{align}

\begin{align}
m = 0.3796 + 1.485 \omega - 0.1644 \omega^2+ 0.01667 \omega^3, for 0.1<\omega<2.0.
\end{align}

Step 2. Use the Wilson equation to calculate an initial guess of $K$ and then determine $\beta$ with the Rachford-Rice procedure (i.e. by solving the Rachford-Rice equation, Eq. (5)):

\begin{align}
K_i^{Wilson} = \frac{p_{c,i}}{p} \exp \left( 5.37(1+\omega_i) \left(1-\frac{T_{c,i}}{T}\right) \right).
\end{align}

Step 3. Solve the following cubic equation to determine the compressibility factor, $Z$, based on the calculation of $A$ and $B$. Thus, fugacities for each phase can be estimated as Equation (3) and (4).

\begin{align}
Z^3 - (1-B)\ Z^2 + (A-2B-3B^2)\ Z\ -(AB-B^2-B^3) = 0, where A = \frac{a(T) p}{ R^2 T^2}, \,\,\, B = \frac{b p}{RT}.
\end{align}
 
Step 4. Check the equilibrium statement, which is often conducted with a convergence test: 

\begin{align}
\left| \frac{f_i^V}{f_i^L}-1 \right| < \epsilon,
\end{align}

where $\epsilon$ is the criterion.   

Step 5. Update $K$ if the criterion is not satisfied (the equilibrium state is not reached), and repeat steps 3 and 4 until the equilibrium has been reached. 

\begin{align}
K_i^{new} = \frac{f_i^L}{f_i^V}K_i^{old},
\end{align}
where $\varphi^V_{i} = \frac{f^V_{i}}{y_i p}$ and $\varphi^L_{i} = \frac{f^L_{i}}{x_i p}$. 

It should be noted that in Step 3, two separate solutions of the cubic equation need to be conducted to calculate $Z$ and then determine the phase molar volumes $v^L$ and $v^{V}$. Obviously this procedure for binary components is more difficult and complex compared to pure fluid flash calculation, as we meet six roots (three for vapor and three for liquid). Further process should be considered for the six roots, for example, the middle roots for vapor/liquid are discarded because they lead to unstable phases. Besides, the remaining roots need to be paired to calculate component fugacities, and the paring selection is a challenge as there are two roots for each phase. If the selection of paring is wrong, the whole procedure will fail with an unstable or metastable solution. Some experiences can be referred to help make the pairing. For example, total Gibbs energy should be minimized for the correct equilibrium solution and sometimes the correct root for liquid should minimize the molar volume. However, for most cases, the best root for vapor phase will maximize the molar volume. }

These restrictions make the original SSM complex and unsteady. Sometimes the roots even fail to present reasonable physical meanings and the root calculation often costs a significant portion of the total CPU time. As a result, an increasing number of studies have been carried out to optimize this method, including the Newton's method, which is also based on PR-EOS and Sparse Grids method.

\subsection{Newton's Method}
{\color{black} Newton-Raphson method, also known as Newton's method, is an optimization procedure to successively estimate the roots of a real-valued function: $f(x) = 0.$. Starting with a good initial guess,  $x_0$ for $f(x) = 0$, a repeating updating will be conducted until a sufficiently accurate value is reached}:

\begin{align}
x_{n+1}=x_{n}-{\frac {f(x_{n})}{f'(x_{n})}}.
\end{align}

The Newton's method can be summarized into 5 steps:

Step 1: Let $n=0$ to obtain initial $K_i$ from SSM, or the previous time step in simulations, etc. 

Step 2: Given $K_i$, the initial estimate of $\beta$ is obtained from the solution of the Rachford-Rice equation:

\begin{align}
F_{M+1}(K_1,K_2,\cdots,K_M,\beta) = \sum_{i=1}^M \frac{(K_i-1)z_i}{1+\beta(K_i-1)}=0.
\end{align}

Step 3: Solve the following linear system to get $\mathbf{x}^{(n+1)}$:

\begin{align}
J_{F}(\mathbf{x}^{(n)})\left( \mathbf{x}^{(n+1)} - \mathbf{x}^{(n)}  \right) = - \mathbf{F}(\mathbf{x}^{(n)}), 
\end{align}

where

\begin{align}
\mathbf{x}^{(n)} = \left( \begin{array}{c}  K_1^{(n)} \\ K_2^{(n)} \\ \vdots \\ K_M^{(n)} \\ \beta^{(n)} \end{array}  \right), \qquad \mathbf{F}(\mathbf{x}^{(n)}) = \left( \begin{array}{c}  F_1(\mathbf{x}^{(n)}) \\ F_2(\mathbf{x}^{(n)}) \\ \vdots \\ F_M(\mathbf{x}^{(n)}) \\ F_{RR}(\mathbf{x}^{(n)}) \end{array}  \right),
\end{align}

\begin{align}
F_i(K_1,K_2,\cdots,K_M,\beta) = \ln K_i - \ln \varphi^L_i + \ln \varphi^V_i = 0,
\end{align}

\begin{align}
F_{RR} = \sum_{i=1}^M y_i - \sum_{i=1}^M x_i.
\end{align}

Step 4: Let $n \leftarrow n+1$.

Step 5: Repeating step 3 and 4 until a sufficient accuracy is reached.

\subsection{Sparse Grids Method}
{\color{black} Various deformations have been proposed in sparse grids methods, while the main idea remains the same \citep{Wu:2015ab}. The approximation of a function is calculated through a summation of a suitable set of basis functions. The basis functions are computed on the set of grid points where the original function is evaluated. For the application in flash calculation, the sparse grids method is split into two steps: online phase and offline phase. A surrogate model is created during the offline phase, and in the online phase it can be cheaply evaluated for the compositional flow simulation.  The biggest advantage of this method is the repeatable usage of the surrogate model, so that only one determination of this model is needed, which greatly saves the computation effort. Generally, the CPU time used for the offline phase (model creation) can be neglected.

Fourteen values are calculated from the original binary component flash calculation, and these values are called observables for the two components in the sparse grids method. To construct a surrogate model, twelve of these values are taken as the input values, while the remaining two observables, liquid and gas, are not considered in the surrogate model. All the fourteen values are presented in Table \ref{tab:observables}. 

Obviously, the simplest and most straightforward approach for creating a surrogate model is to store all the evaluations resulted from original flash calculations as listed in the above table on a regular Cartesian grid. The problem is that this full grid surrogate can result in out-of-memory for large systems with high order of components. Besides, in fact, not all the grids are used during calculation, which causes the waste of memory.

So we move to the sparse grids method, which reduces the requirements of memory. The 12 adaptive sparse grid approximations $S_{s,i}$ with  $i=1,\ldots,12$ are constructed. During the offline phase, each $S_{s,i}$ starts with a basic low resolution sparse grid which is then refined in the areas with the highest surplus. Since each flash calculation retrieves all the 12 output values, the construction of an $S_{s,i}$ can reuse values if a particular grid point, i.e. a parameter combination $(p,z_{1})$, has already been evaluated by a flash calculation for building another $S_{s,i}$. Using this method, the computation effort to create the surrogate model is largely reduced. It has been reported that, because all the observables can be calculated through the original flash calculation, the union of all the grid points is sometimes used as the union sparse grids for each of the surrogates $S_{s,i}$, which increases significantly the data stored in memory but only a minor approximation accuracy will be obtained. Thus, it is not a good idea to use the union sparse grids. It is also noted that due to the basis functions overlapping, $S_{s}$ will always present a slightly larger overhead evaluation compared with $S_{f}$ ($f$ for full grid surrogate model), but this overhead could be neglected. }


\begin{table}
\caption{Observables in sparse grids methods for flash calculations}
\label{tab:observables}

\begin{center}
\begin{tabular}{|c|c|}
\hline 
Observable & Physical quantity\tabularnewline
\hline 
\hline 
$xW1$ & Molar fraction of Component 1 in the oil phase\tabularnewline
\hline 
$xW2$ & Molar fraction of Component 2 in the oil phase\tabularnewline
\hline 
$xN1$ & Molar fraction of Component 1 in the gas phase\tabularnewline
\hline 
$xN2$ & Molar fraction of Component 2 in the gas phase\tabularnewline
\hline 
$xiW$ & Molar density of oil phase (unit: $mole/m^{2}$)\tabularnewline
\hline 
$xiN$ & Molar density of gas phase (unit: $mole/m^{2}$)\tabularnewline
\hline 
$densiW$ & Mass density of oil phase (unit: $kg/m^{2}$)\tabularnewline
\hline 
$densiN$ & Mass density of gas phase (unit: $kg/m^{2}$)\tabularnewline
\hline 
$sW$ & Saturation\tabularnewline
\hline 
$v1$ & Partial molar volume of Component 1\tabularnewline
\hline 
$v2$ & Partial molar volume of Component 2\tabularnewline
\hline 
$Cf$ & Isothermal compressibility of the flow\tabularnewline
\hline 
$Liquid$ & Boolean variables, 1 for liquid.\tabularnewline
\hline 
$Gas$ & Boolean variables, 1 for gas.\tabularnewline
\hline 
\end{tabular}
\par\end{center}
\end{table}

\section{Artificial Neural Networks}
\begin{figure}[htbp]
\centerline{\includegraphics[width = 100mm]{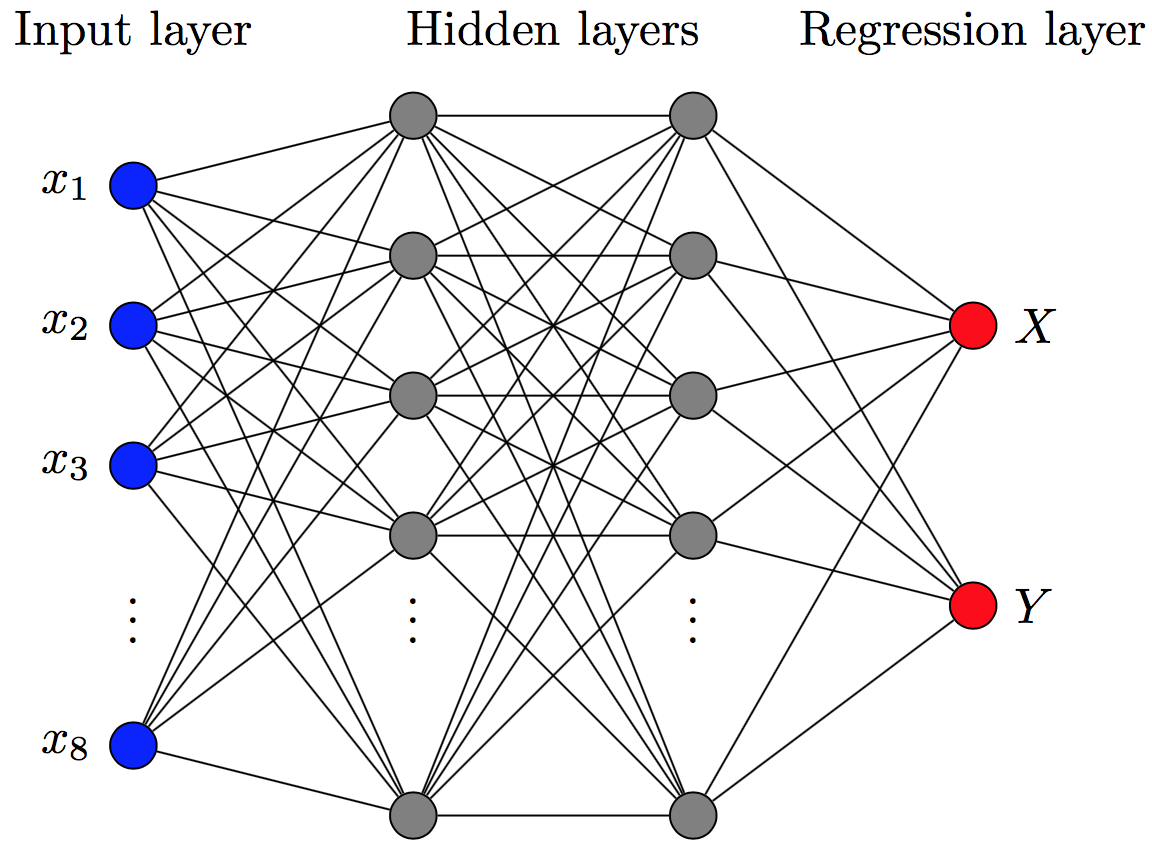}}
\caption{The neural network to model VLE.} \label{fig_network1}
\end{figure}

Artificial neural networks (ANNs) are computational models designed to incorporate and extract key features of the original inputs.
Deep neural networks usually refer to those artificial neural networks which consist of multiple hidden layers.
As shown in Figure \ref{fig_network1}, a deep fully connected neural network is applied to model the VLE. Following the input layer, a number of fully connected hidden layers, with a certain number of nodes, stack over the other, whose final output is fed into another fully connected connected layer, which is the final output layer. Since we are fitting X and Y in our model, the final output layer contains two nodes, each of which predicts the value of one of the two variables. The activation function of this layer is fixed as linear. Naturally the proposed ANN input variables include critical
pressure ($Pc$), critical temperature ($Tc$) and acentric factor ($\omega$) of the components comprising the mixture. As a result, the eight variables in Figure \ref{fig_network1} are the above three factors for each of the two components in the mixture, the temperature, and the pressure. 
The required $C_{1}$
to $C_{7}$ binary mixture experimental VLE data were gathered from the Korea Thermophysical Properties Data Bank (KDB), of 1332
data points in total, with supplementary selection of consistency and applicability. 
As instructed on the database, the expected mean relative error of the experimental data we used for training and validating the model is around $20\%$.  
A large range of pressures and temperatures are considered while ensuring that the mixture does not enter into a critical state, which is to confirm that a two-phase condition is ensured. 

\begin{figure}[htbp]
\centerline{\includegraphics[width = 90mm]{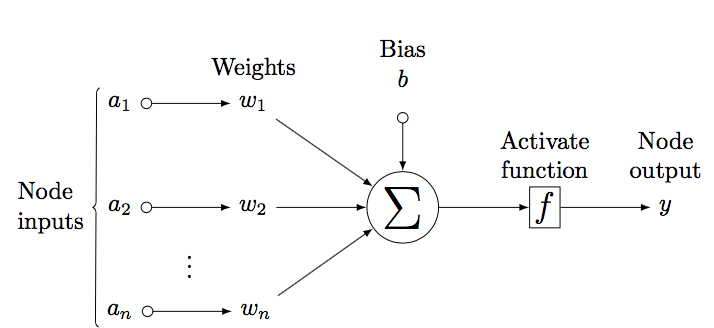}}
\caption{The flow chart of the simulation process in each node} \label{fig_network2}
\end{figure}

Due to the high complexity of the neural network model and the limited number of data (only 1332 records in total), the trained model is subject to overfitting. To deal with the common and most serious issue in the deep learning field, we adopted weight decay as well as dropout \citep{RN21} to handle the problem. The model initialization can also influence the final result significantly. We utilized Xavier initializer to perform the model initialization. The whole package is developed using TFlearn. Trained on a workstation with one Maxwell Titan X card, the model converged in 10 minutes. A simplified flow chart of our network model working process in each node is presented in Figure \ref{fig_network2}.

Formally, for the $i$-th hidden layer, let $\textbf{a}_i$ denote the input of the layer, and $\textbf{y}_i$ to denote the output of the layer. Then we have:
\begin{equation}
\textbf{y}_i = \textit{f}_i(\textbf{W}_i*\textbf{a}_i+\textbf{b}_i),
\label{eq:layer}
\end{equation}
where $\textbf{W}_i$ is the weight; $\textbf{b}_i$ is the bias; and $\textit{f}_i$ is the activation functions of the $i$-th layer. For a network with multiple layers, the output of one hidden layer is the input of the next layer. For example, we can represent the network in Figure \ref{fig_network1} as:
\begin{equation}
\textbf{o} = \textit{f}_3(\textbf{W}_3*\textit{f}_2(\textbf{W}_2*\textit{f}_1(\textbf{W}_1*\textbf{x}_1+\textbf{b}_1)+\textbf{b}_2)+\textbf{b}_3),
\label{eq:net}
\end{equation}
where $\textbf{o} = (X, T)$; $\textit{f}_1, \textit{f}_2, \textit{f}_3$ are the activation functions; $\textbf{W}_1, \textbf{W}_2, \textbf{W}_3$ are the weights for each layer; $\textbf{b}_1, \textbf{b}_2, \textbf{b}_3$ are the bias terms of each layer.

Here are the short explanations of the techniques used to obtain a practical network:
\begin{enumerate}
  \item Weight decay: Overfitting is usually a serious issue in the deep learning field, which means that the learned model has almost perfect performance on the training data while performs poorly on the validation or testing data. The main reason of overfitting in this field is that the model itself is composed of too many parameters while we do not have enough training data, that is, the model is over-parameterized. In order to prevent the overfitting issue from hurting the model's performance, we usually apply additional constraint on the model's parameters to reduce the freedom of the model. In general, if the model is overfitted, the norm of the weight parameters is often very large. As a result, one way to avoid overfitting is to add an additional constraint on the norm of the weight parameters and penalize large weights. In practice, we can add a regularization term, which is related to the norm of the weights, in the loss function to make the model fit the training data and penalize large weights at the same time. Formally, the original loss function for deep learning, which is the mean squared loss in our problem, can be formulated as:
  \begin{equation}
  L = \frac{1}{N}\sum_{n=1}^{N}\norm{\textbf{o} - \hat{\textbf{o}}}^2,
  \label{eq:loss}
  \end{equation}
  where $N$ is the total number of training data; $\bf{o}$ is the output of the model; $\hat{\bf{o}}$ is the observed value.
  After adding the L2 weight decay term, the loss function becomes:
  \begin{equation}
  L = \frac{1}{N}\sum_{n=1}^{N}\norm{\textbf{o} - \hat{\textbf{o}}}^2 + \lambda \norm{\bf{W}}^2_{2},
  \label{eq:loss_reg}
  \end{equation}
  where $\bf{W}$ is the whole set of weight parameters of the model; $\lambda$ is the regularization coefficient, i.e., how much we penalize over the large weights.

  \item Dropout: Dropout is a very efficient method for dealing with overfitting in neural networks. This method reduces the freedom of the network by discarding nodes and connections of the model during the training stage. For example, if we apply the dropout technique to a certain layer with the keep probability as $p$ $(0<p<1)$, then, during each training stage, each node of that layer would first be evaluated independently with the probability of $p$ being kept or the probability of $1-p$ being discarded. If the nodes are discarded, all the nodes and connections are discarded from the model. After the dropout procedure, the reduced network is trained during the training stage. After that certain training stage, the discarded nodes are inserted to the model with the original weights and the model enters the next training cycle. 

  \item Xavier initializer: The initialization of the neural network model is of vital importance, which can affect the convergence speed and even the final model's performance. If the weights are initialized with very small values, the variance of the input signal vanishes across different layers and eventually drops to a very low value, which reduces the model complexity and may hurt the model's performance. If the weights are initialized with very large values, the variance of the input signal tends to increase rapidly across different layers. That may cause gradient vanishing or explosion, which increases the difficulty of training a working model. Since we usually initialize the weights with a Gaussian distribution, to control the variance of the signal, it is desirable to initialize the weights with a variance $\delta$ to make the variance of the output of a layer the same as that of the input of the layer. 
  For example, for Figure \ref{fig_network2}, we want:
  \begin{align}
  var(y) & = var(w_1*a_1+w_2*a_2+...+w_n*a_n+b) \nonumber \\
    & = var(w_1)*var(a_1) + var(w_2)*var(a_2) +...+ var(w_n)*var(a_n) \nonumber \\
    & \stackrel{(1)}{=} n*var(w_i)*var(a_i), 
  \end{align}
  where (1) is because of the identity distribution assumption of all the $w_i$ and $a_i$. Since we want the variance of $y$ the same as the variance of $a_i$, we need:
  \begin{align}
  n*var(w_i) = 1,
  \end{align}
  as a result, we should initialize the weight of each layer using Gaussian distribution with the variance as $\frac{1}{n}$, where n is the number of weights in that layer. This initializer is known as the Xavier initializer.
  \item Batch normalization: Training deep learning model is notoriously time-consuming, because of the large number of parameters belonging to different layers. Not only is the optimization for such a large number of parameters internally time-consuming, but there are some undesirable properties of the multi-layer model which makes the convergence process slow. One property of the deep learning method is that the distribution of each layer's input might change because the parameters of the previous layer are usually changed during training, which is usually referred to as ``internal covariate shift''. To solve the problem, batch normalization is proposed. In addition to normalize the original input of the model, which is the input of the first layer, this technique makes the normalization part of the model and performs normalization on hidden layers for each training batch during the training stage. Batch normalization enables larger learning rates and can accelerate the convergence speed by 10 times.

  \item Activation functions: As shown in Equation (\ref{eq:net}), the activation function is where the non-linearity and the expressiveness power of deep neural network models comes from. There are numerous activation functions: Rectified linear unit (ReLU), Parameteric rectified linear unit (PReLU), TanH, Sigmoid, Softplus, Softsign, Leaky rectified linear unit (Leaky ReLU), Exponential linear unit (ELU), and Scaled exponential linear unit (SELU). The most commonly used ones are ReLU 
  \begin{equation}
  f(x) = 
  \begin{cases}
    0,        & \text{if } x<0\\
  x,      & \text{if } x \geq 0,
  \end{cases}
  \end{equation}

  and Sigmoid:
  \begin{equation}
  \sigma(x) = \frac{1}{1+\exp(-x)}.
  \end{equation}
  In Section \ref{sec:Section-title}, we show the performance of our model with different activation functions.

\end{enumerate}

\section{Results and Discussion\label{sec:Section-title}}

In this paper, to accelerate and optimize the original flash calculation using SSM, attempt has been made to use the deep learning method for the Vapour-Liquid equilibrium (VLE) calculation of the systems $C_{1}-C_{7}$ mixtures, including Methane, Ethane, Propane, N-Butane, N-Pentane, N-Hexane, N-Heptane in NPT kind flash. Two other accelerating methods, Newton's Method and Sparse Grids Method, are also introduced and used as a comparison. Physical properties of each component are listed in Table \ref{tab:property}. Essentially, as the Gibbs phase rule stipulates, two
intensive properties are required to completely describe a binary two-phase system at equilibrium conditions. Temperature and pressure are two such thermodynamic intensive properties conventionally selected, because of the relative ease with which they can be measured. Alongside temperature and pressure, the acentric factor is also generally included in VLE phase equilibrium calculations to account for non-sphericity of molecules. The required C1 to C7 binary mixture experimental
VLE data were gathered from the Korea Thermophysical Properties Data Bank (KDB), of totaling 1332 data points, with supplementary selection of consistency and applicability. As instructed on the database, the expected mean relative error of the experimental data we used for training and validating the
model is around $20\%$. A large range of pressures and temperatures are considered while ensuring that the mixture does not enter into a critical state, which is to confirm that a two-phase condition is ensured.

\begin{table}
\caption{Physical properties of components involved in this
paper}
\begin{center}
\begin{tabular}{|c|c|c|c|}
\hline 
Component & $T_{c}(K)$ & $P_{c}(bar)$ & $\omega$\tabularnewline
\hline 
\hline 
$C_{1}$ & 190.6 & 46 & 0.0115\tabularnewline
\hline 
$C_{2}$ & 305.4 & 48.84 & 0.0908\tabularnewline
\hline 
$C_{3}$ & 369.8 & 42.46 & 0.1454\tabularnewline
\hline 
$C_{4}$ & 421.09 & 37.69 & 0.1886\tabularnewline
\hline 
$C_{5}$ & 467.85 & 34.24 & 0.2257\tabularnewline
\hline 
$C_{6}$ & 521.99 & 34.66 & 0.2564\tabularnewline
\hline 
$C_{7}$ & 557.09 & 32.62 & 0.2854\tabularnewline
\hline 
\end{tabular}
\par\end{center}
\label{tab:property}
\end{table}

\begin{figure}[t]
\centerline{\includegraphics[width = 70mm]{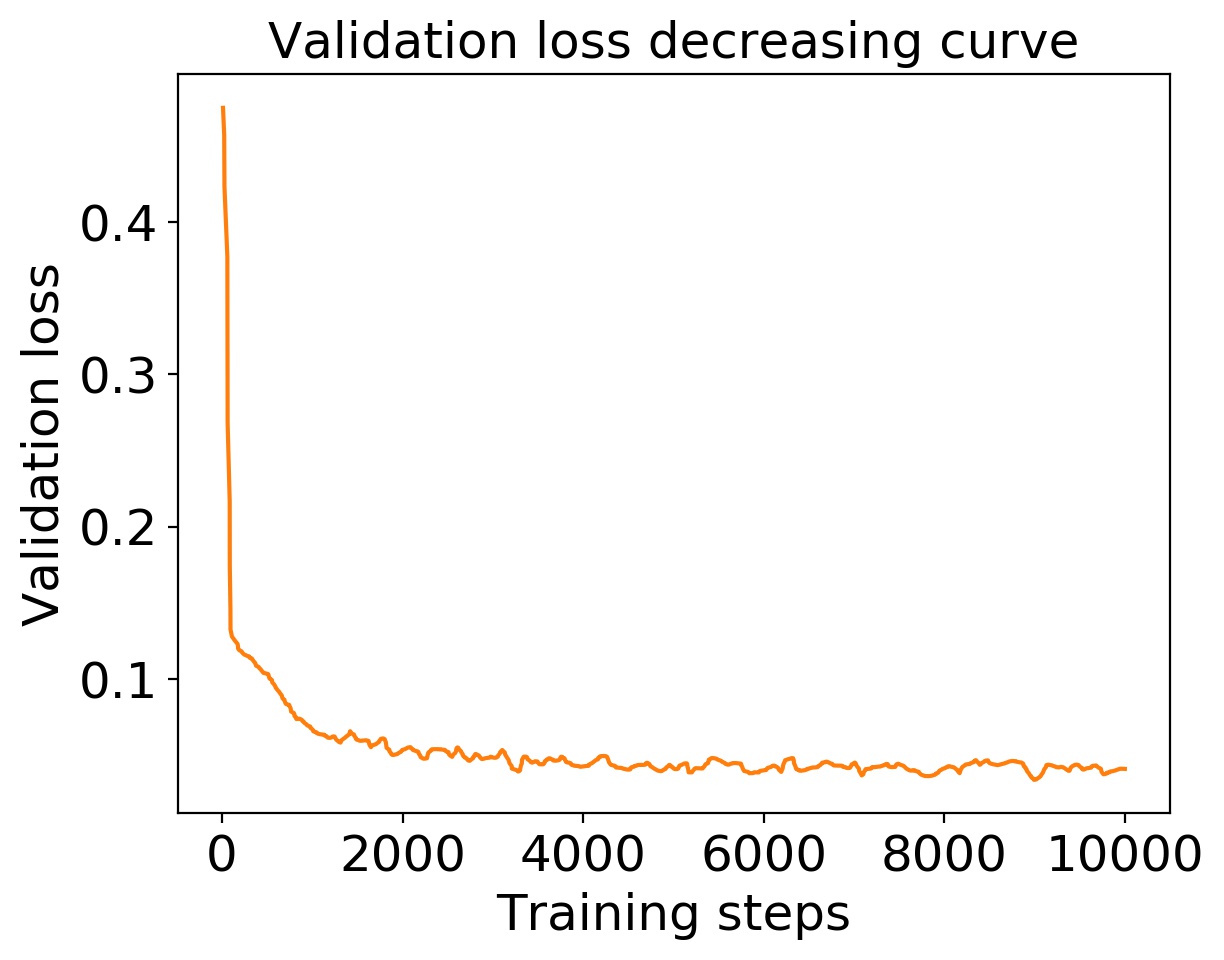}}
\caption{The validation loss decreasing curve.} 
\label{fig:loss}
\end{figure}
{\color{black}
\subsection{Training a Deep Learning Model}
In this section, we briefly describe how we train a neural network on this problem. We use a neural network with the following configurations: the neural network has 5 layers, with 100 nodes within each layer and ELU as the activation function. The performance of the neural networks with other configurations will be discussed in Section \ref{sub:factors}. As shown in Figure \ref{fig_network1} and Figure \ref{fig_network2}, the model takes the eight variables as input and outputs the predicted value of $X$ and $Y$. The key parameters of the model are the weights of each layer, which control what the model outputs given the input. At first, those weights are initialized randomly, which means that the model will output useless values given the inputs. To make the model useful for this problem, we need to train those weight parameters to fit our problem. In doing so, we run an iterative optimization process on the model's parameters to make the difference between the model's output and expected value (ground truth) on the training data as small as possible. The difference between the model's output and the ground truth is referred as loss. Here, for this regression problem, we use mean square error as the loss function, which has the following form:
\begin{equation}
L = \frac{1}{N}\sum^N_{n=1}[(X_n-\tilde{X}_n)^2+(Y_n-\tilde{Y}_n)^2],
\end{equation}
where $X_n, Y_n$ are the outputs of the deep learning model and $\tilde{X}_n, \tilde{Y}_n$ are the ground truth of data point $n$; N is the total number of training data points. Since we have 1332 data points in total, we use 1199 data points as the training data and 133 data points as the validation data, which is not used in the optimization process but used as a separate dataset to oversee the model's performance. We use Adam \cite{RN102} as the optimizer. During each step, we feed the model with a batch of 256 data points randomly selected from the training data. We show the training process in Fig. \ref{fig:loss}. As shown in the figure, with the training going on, the loss of the model on the validation dataset decreases gradually, which means the difference between the model's output and the expected value is becoming smaller, and the model becomes increasingly useful.
}

\subsection{Effects of Factors in Deep Learning Method}
\label{sub:factors}
\begin{figure}[htbp]
\centerline{\includegraphics[width = 140mm]{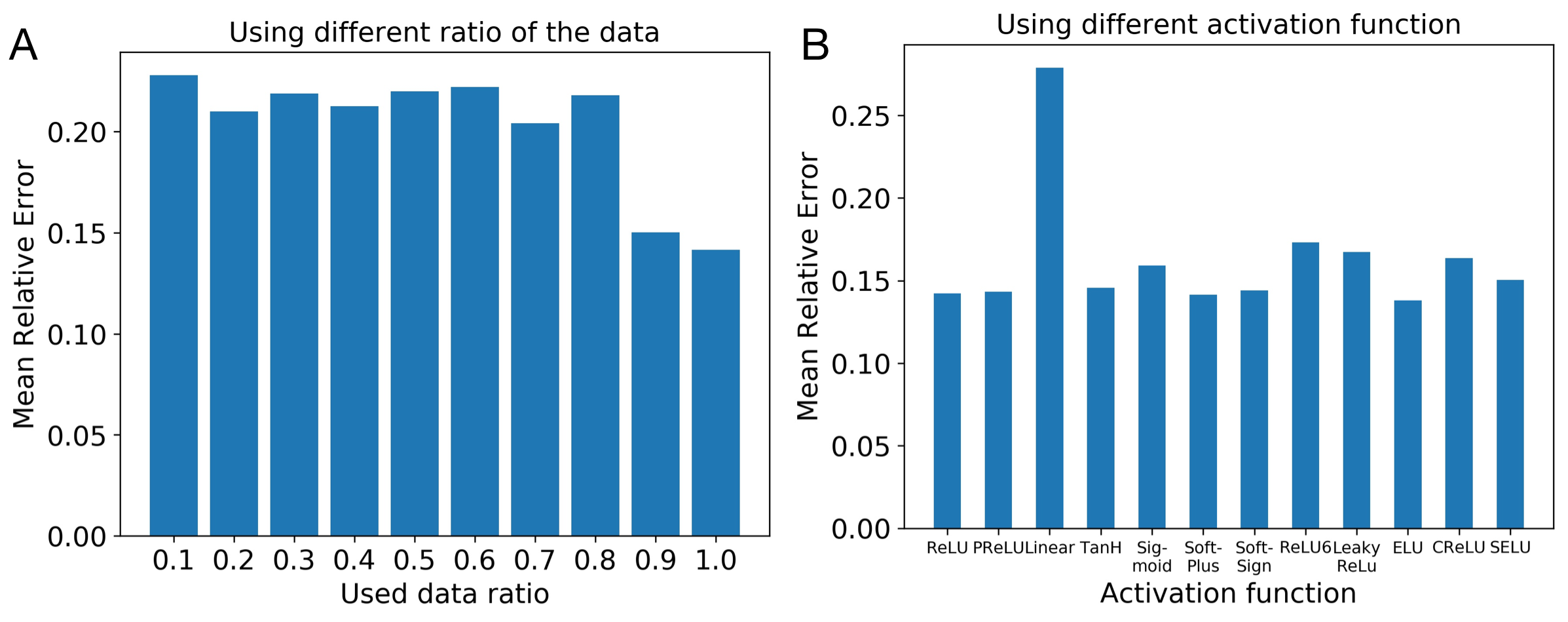}}
\caption{(A) The relationship between the deep neural network's performance and the data size we used. We fixed the activation function as RELU and the number of hidden layers as 3, with 200 nodes in each layer. (B) The performance comparison of the deep neural network using different activation functions. We fixed the number of hidden layers as 3 and the number of nodes in each hidden layer as 200.} 
\label{fig_size}
\end{figure}

To find a more comprehensive neural network, especially suitable for VLE prediction, we investigate the effects of different factors in the model. It can be referred in Figure \ref{fig_size} (A) that the mean error of our trained model generally performs a slight correlation with the data size, at around $20\%$ with used data ratio varying from $10\%$ to $80\%$, while decreases significantly if this ratio reaches up to $90\%$. As the data itself is noisy, with a mean $20\%$ relative error, this result is acceptable without the risk of overfitting. 

Besides, twelve different activation functions under 3 hidden layers and 200 nodes at each layer are compared. The results are shown in Figure \ref{fig_size} (B). It can be inferred that the linear function performs relatively worse, with the mean relative error higher than 0.25. It is interesting to see that other functions will result in a generally similar mean error. Specifically, other functions can be divided into two groups: one with lower errors, including ReLU, PReLU, TanH, SoftPlus, SoftSign and ELU, while the other group consists of Sigmoid, ReLU6, Leaky ReLU, CReLU and SELU. A clear boundary dividing these two is the mean error of 0.15. In this paper, the RELU function in the lower group is selected as the activation function. 

\begin{figure}[htbp]
\centerline{\includegraphics[width = 140mm]{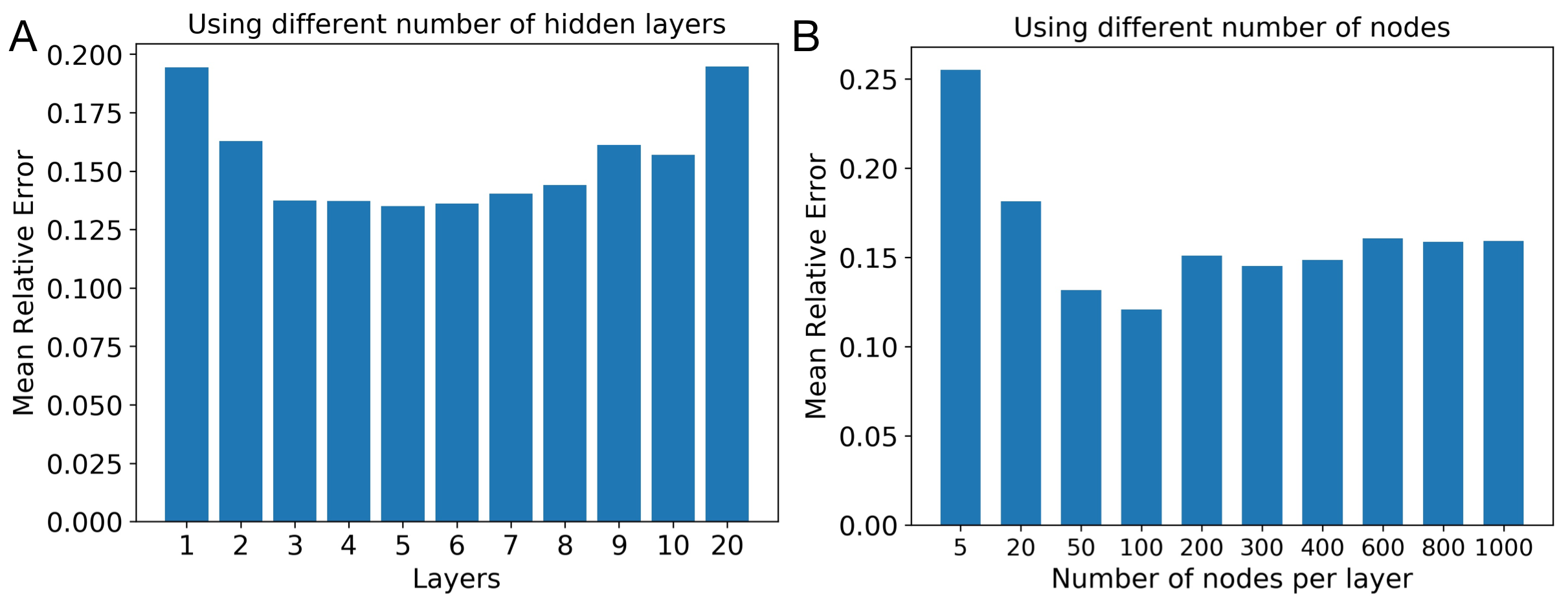}}
\caption{(A) The performance comparison of the deep neural network using different number of hidden layers. We fixed the number of nodes in each hidden layer as 200 and employ the RELU activation functions. (B) The performance comparison of the deep neural network using different number of nodes in each of the hidden layers. Following the previous experiment, we fixed the number of layers as 5 and the activation function as ELU.} 
\label{fig_layer}
\end{figure}

Another interesting finding is that more layers cannot ensure a lower mean error in this model trained for phase equilibrium prediction. As shown in Figure \ref{fig_layer} (A), there is an obvious decreasing trend in mean relative error when the layer number increases from 1 to 3. For 3-6 layers, only a very slight difference can be seen in the main relative error. It is surprising to see that the error will increase with the raising of layer numbers for more than 7. For 20 layers, which is truly a very complex system, the mean relative error is as high as that of only 1 layer. It can be concluded that 3-6 hidden layers are better for the prediction, and in this paper we should consider both the efficiency and accuracy. 

We further compare the performance of our model with different number of nodes per layer. The results can be seen in Figure \ref{fig_layer} (B). It is reasonable that the mean relative error will decrease if the number of nodes increases from 5 to 100, and then remains nearly the same from 200 to 1000. To our common knowledge, more complex models with more nodes will perform better, as long as we can regulate the model well during training, with certain extreme as we are using a noisy database. The test case proves that.

\subsection{Comparison of Different Methods in Flash Calculation}

A selected model based on the above analysis has been used in a flash calculation case, and the results are compared with three other methods introduced in Section 2, Successive Substitution Method, Newton's Method and Sparse Grids Method. The binary components in this case are set as Methane and Propane, with temperature constant at 226K and pressure of 11 values changing from 6bar to 77bar. The CPU time used for each method is listed in Table \ref{tab:cpu}. 

\begin{table}
\caption{CPU time used in different methods}
\label{tab:cpu}
\begin{center}
\begin{tabular}{|c|c|c|}
\hline 
Results source & CPU time (s) & Acceleration\tabularnewline
\hline 
\hline 
SSM & 2503.32 & 1 \tabularnewline
\hline 
Newton & 1201.76 & 2.082\tabularnewline
\hline 
Sparse grids & 5.11 & 489.823\tabularnewline
\hline 
Deep Learning & 1.22 & 2051.639 \tabularnewline
\hline 
\end{tabular}
\end{center}
\end{table}

It should be noted that the initial guess of Newton's method is the result of SSM, which means that it will take much less time to converge. Besides, the time used to generate the surrogate model in the sparse grids method is neglected, as treated in \cite{Wu:2015ab}, which means that the total real CPU time for sparse grids are much higher. In fact, SSM is applied to get the initial data model. For the deep learning model, the CPU time for data training is also neglected, but this training time is only 15.72 seconds, which is much lower than other ones. Meanwhile, the trained model can be repeatedly used for different binary components and the conditions of temperature and pressure. It can be concluded that the deep learning method is much more efficient than the traditional SSM, and also faster than the other two acceleration methods. It is easy to expect better efficiency of Sparse Grids and Deep Learning method in large scale calculation, as the model of the two can be repeatedly used in different cases, but in SSM and Newton's method everything will start from scratch. 

\begin{figure}[htbp]
\centerline{\includegraphics[width = 180mm]{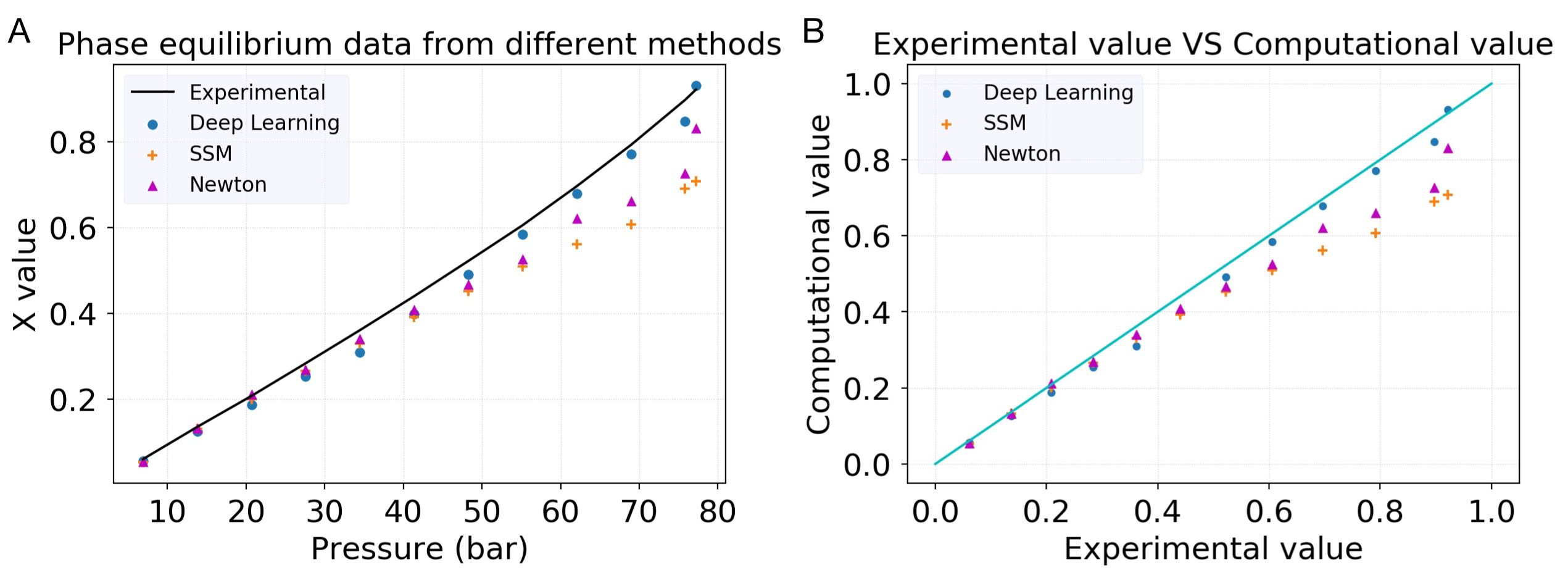}}
\caption{The performance comparison of different calculation methods and experimental data. (A) The relationship between the pressure and four kinds of X values (experimental values and three computationally calculated values). (B) Direct comparison between the experimental data and the three computationally calculated data. The closer the data points is to the diagonal, the better the computational method is in terms of calculation accuracy.} 
\label{fig_comp}
\end{figure}

Except for efficiency, the accuracy of our optimized deep learning model is also proved in our calculation. We collected the experimental data at certain temperature, composition and pressure conditions as the ground truth and compared with the results from different calculation methods. It is noted that as the sparse grids method is based on a surrogate model generated from SSM, the results are neglected in the comparison. It can be referred from Figure \ref{fig_comp} that all the results calculated from these methods match the experimental data well, although not perfectly. Generally speaking, SSM will show obvious error at some points, but results from Newton's method are much better as they converge from the result of SSM. The results of Newton's method is much better, but still less accurate than Deep Learning model. In summary, the optimized deep learning model has much better CPU time efficiency while conserving the similar accuracy of other flash calculation methods.  

\subsection{Comparison of Prediction and Experimental Data}
\begin{figure}[htbp]
\centerline{\includegraphics[width = 140mm]{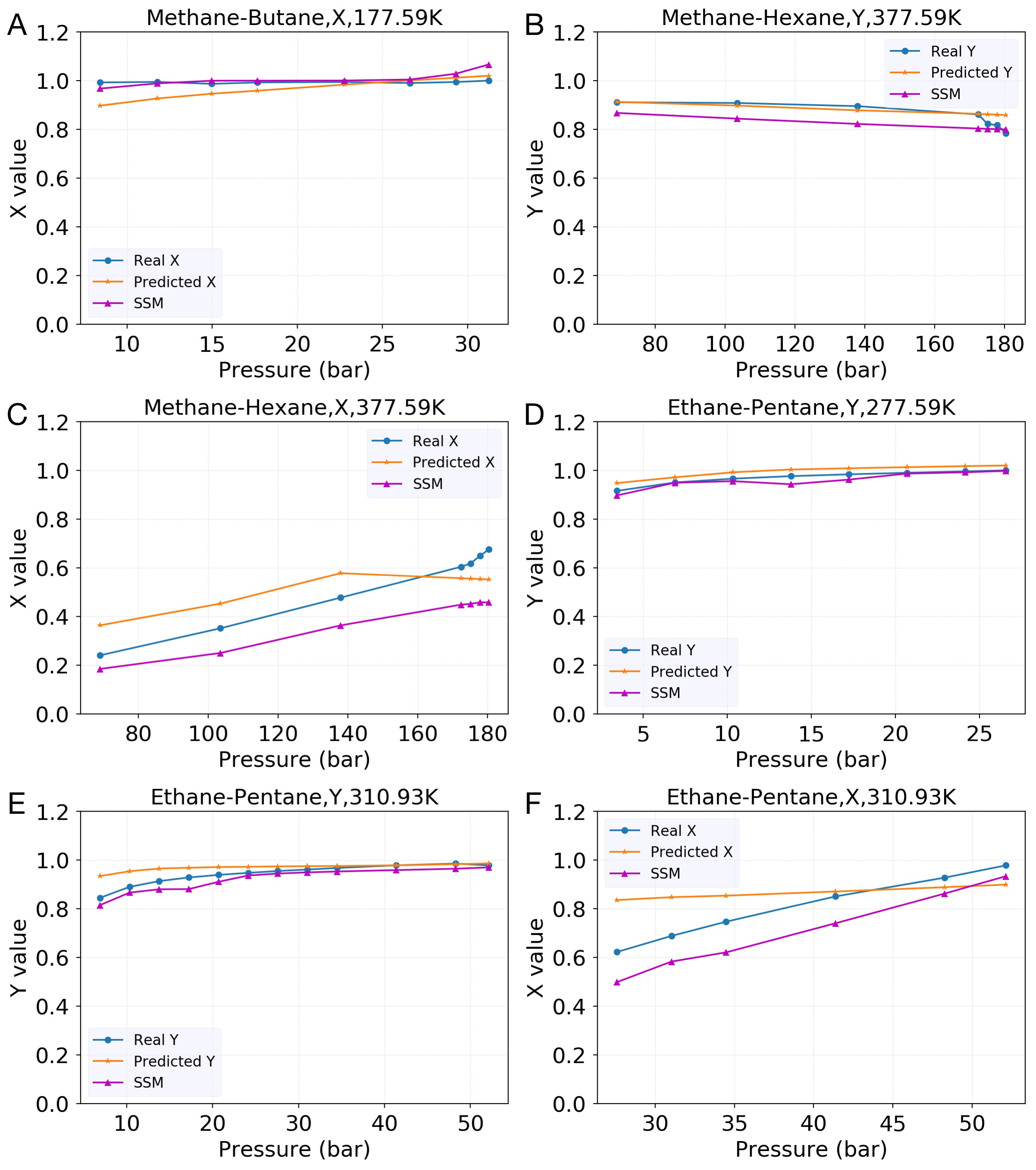}}
\caption{VLE prediction with experimental data.} 
\label{fig_res}                                                        
\end{figure}

To demonstrate the relative accuracy of our proposed model with experimental data (seen as real data), Figure \ref{fig_res} illustrates six sample isothermal VLE diagrams of different binary component combinations. The $X$ value represents the mole fraction of the liquid phase of the first component and the $Y$ value represents that of the vapor phase. All the cases are isothermal VLE processes and x-label in the diagrams represents the pressure. It is indicated that the most significant shortcomings of the conventional VLE calculation procedures based on various EOSs, known as the disability to guarantee convergence to the intended results, are overcomed by our deep learning model. Both the vapor and liquid mole fractions predicted match well with the experimental data, notably the availability to handle different VLE processes with variables of component and temperature conditions. In the presented cases, the highest relative error is found as about $25\%$ ,while in some cases, like Figure \ref{fig_res}(B) and Figure \ref{fig_res}(D), the relative error is less than $5\%$. As the experimental data has its own relative error of $20\%$ , these results can be seen as a verification of the reliability to our proposed models. 

More details can be investigated from the six samples. It can be inferred from Figure \ref{fig_res}(A) and Figure \ref{fig_res}(B) that for different component and temperature conditions, our model will result in different compatibility. However, from Figure \ref{fig_res}(D) and Figure \ref{fig_res}(E), it is easy to find that for the same binary components, the performance of our proposed model varies at different temperatures. For $Y$, the vapor mole fraction, it seems that the prediction error increases with the temperature. From the comparison of Figure \ref{fig_res}(B) and Figure \ref{fig_res}(C), as well as the comparison of Figure \ref{fig_res}(E) and Figure \ref{fig_res}(F), it can be found that for the same combination of binary components and temperature, the prediction error varies as well for liquid and vapor mole fraction. In general, prediction of the vapor mole fraction using our model performs better than the liquid mole fraction, while the latter prediction has a much larger relative error. 

To show the advantage of our deep learning model on VLE estimation, Successive Dubstitution Method (SSM) has also been performed for each flash calculation problem as a comparison. Generally speaking, it is indicated that the VLE calculation using Deep Learning Method can always preserve the accuracy, while SSM may result in some large error, like shown in Figure \ref{fig_res}(F). However, for some cases, like shown in Figure \ref{fig_res}(A), the original SSM flash calculation performs better at low pressure. In all, it could be stated that model trained from deep learning has better stability, not accuracy, compared with traditional flash calculation method. Besides, it need to be pointed out that in the SSM calculation, the selection of roots pairing are done manually, which effects the computation efficiency and the whole process repeatability. 

\section{Conclusion\label{sec:Section-title-1}}

In this study, the capacity of the deep learning method in making fast and reliable predictions of equilibrium compositions of alkane binary mixtures is investigated. A data set comprising 1332 data points was gathered and used in training the proposed ANN model and testing. The results obtained demonstrate the relative precision of the proposed deep learning model and process, and the capability of ensuring a converged result of VLE. Compared with traditional SSM and two other acceleration methods, Newton's method and Sparse Grids method, our optimized Deep Learning model performs more efficiently while conserving similar accuracy. Furthermore, we make some remarks based on the results:

\subparagraph{1}

Concisely, the proposed models can serve the
purpose of being close first estimates for more thermodynamically rigorous vapor–liquid equilibrium calculation procedures. However, overfitting is obvious in the model construction process and results in relatively high prediction errors in some cases. Thus, it is still necessary to construct a larger set of experimental binary VLE data, which is hard to collect based on current database. 

\subparagraph{2}
Another way to get large amount of data is to use flash calculation with EOS. However, it is commonly acknowledged that the convergence and accuracy of current flash calculation methods cannot be ensured, as we have shown in Section 4. Besides, sometimes the results even have no physical meanings and we need to exclude them manually. Thus, the application of flash calculation results used as input data should be studied based on the selection and optimization of the calculation method, which remains to be the future work. Compared with traditional flash calculation method, our deep learning model will show better stability, which means that it can always ensure a reasonable result with acceptable error. The problem of manul handling in the process of flash calculation should also be treated carefully. 

\subparagraph{3}
In our results, it can be concluded that there is no deep learning model perfectly fitting all the VLE conditions. Generally speaking, model performance is dependent on the training data and model parameters. Our proposed model is constructed by the selection of the best factors, including data ratio, active function, layer and node numbers but the prediction error is still relatively high at some cases. Except for more training data, another potential solution is to find a better model combined with the feature of VLE process, especially the characteristics of EOSs. 

\section*{Acknowledgments}
The research reported in this publication was supported in part by funding from King Abdullah University of Science and Technology (KAUST) through the grant BAS/1/1351-01-01 and BAS/1/1624-01-01. 

\bibliographystyle{plain}
\bibliography{references}

\end{document}